\documentclass[aip,jcp,reprint,footinbib,twocolumn,longbibliography]{revtex4-1}
\usepackage{graphicx}
\usepackage{amsmath}
\usepackage{amssymb} %
\usepackage{braket}
\usepackage{float}
\usepackage{bbold}
\usepackage{color}
\usepackage{bm}
\usepackage{subcaption}
\usepackage{comment}
\usepackage{placeins}

\newcommand{\eu}{\mathrm{e}^}

\newcommand{\half}{{\ensuremath{\frac{1}{2}}}}

\newcommand{\Tr}{\text{Tr}}

\newcommand{\Hop}{\hat{H}}

\newcommand{\op}[1]{\ensuremath{\hat{#1}}}

\newcommand{\del}{\nabla}

\DeclareMathOperator{\Real}{Re}
\DeclareMathOperator{\Imag}{Im}
\renewcommand{\Re}{\Real}
\renewcommand{\Im}{\Imag}

\newcommand{\grad}{\del}

\newcommand\identity{1\kern-0.25em\text{l}}

\usepackage{dcolumn} %
\newcolumntype{d}[1]{D{.}{.}{#1}} %

\usepackage{bbold}

\begin{document}
\title{Nonadiabatic ImF instanton rate theory}

\author{Rhiannon A. Zarotiadis}
\email{rhiannon.zarotiadis@nyu.edu}
\affiliation{Department of Chemistry and Applied Biosciences, ETH Z\"{u}rich, 8093 Z\"{u}rich, Switzerland}
\affiliation{Simons Center for Computational Physical Chemistry, New York University, New York, NY 10003, USA}
\affiliation{Department of Chemistry, New York University, New York, NY 10003, USA}
\author{Jeremy O. Richardson}
\email{jeremy.richardson@phys.chem.ethz.ch}
\affiliation{Department of Chemistry and Applied Biosciences, ETH Z\"{u}rich, 8093 Z\"{u}rich, Switzerland}
\date{\today}

\begin{abstract}
Semiclassical instanton theory captures nuclear quantum effects such as tunnelling in chemical reactions.
It was originally derived from two different starting points, the flux correlation function and the ImF premise.
In pursuit of a nonadiabatic rate theory, a number of methods have been proposed; almost all based on the less rigorous ImF premise.
Only recently, we introduced a rigorous nonadiabatic ring-polymer instanton rate theory in the flux-correlation function framework which successfully bridges from the Born--Oppenheimer to the golden-rule limit.
Here, we examine the previous ImF-based attempts and conclude that they do not capture the two limits correctly.
In particular, we will highlight how the last in a series of developments, called mean-field ring-polymer instanton theory, breaks down in the golden-rule limit.
We develop a new nonadiabatic ImF rate theory to remedy the failings of previous attempts while taking inspiration from them.
We also consider the crossover from deep tunnelling to a high-temperature rate theory.
We test our new nonadiabatic ImF theory on a range of models including asymmetric and multidimensional systems and we show reliable results for the deep-tunnelling regime but limitations for the related high-temperature rate theory. 
\end{abstract}

\maketitle 

\section{Introduction}
Reaction mechanisms lie at the heart of much of chemistry, biology and physics from photovoltaic cells to proton-coupled electron transfer in proteins or redox-flow batteries.\cite{HammesSchiffer2015PCET} 
In order to simulate many of these reactions and calculate their rate constants, it is necessary to account for nonadiabatic effects.
Instanton rate theory\cite{Miller1975quantization} is an ideal candidate for studying chemical reactions.
It is derived using the semiclassical approximation to the path-integral formulation of quantum mechanics\cite{Feynman} and captures nuclear quantum effects such as zero-point energy and nuclear tunnelling at a reasonable computational cost.\cite{Perspective}
In recent years, it has also been applied to a variety of molecular systems thus proving a readily-available tool.\cite{Rommel2012enzyme, Asgeirsson2018instanton, broadtop, Heller2021Thiophosgene, Heller2022nitrenes, oxygen, carbenes, porphycene, Hgraphene, methanol, TEMPO, CIinst}
Instanton theory has, until recently, only been derived rigorously in the Born--Oppenheimer (BO)\cite{AdiabaticGreens,InstReview} and golden-rule (GR) limits.\cite{GoldenGreens, GoldenRPI, Heller2020, Richardson2024}
However, many systems of interest find themselves intermediate between these two limits.

In recent work, \cite{Zarotiadis2025} we have developed a rigorous nonadiabatic ring-polymer instanton (NRPI) rate theory applicable to the full range of coupling strengths, and we have shown its successful application in a proof-of-principle study.
In addition to predicting the rate, it provides mechanistic understanding, while remaining of similar computational cost as other instanton rate theories.
Our NRPI theory is derived from first principles in the flux-correlation function framework.\cite{Miller1974QTST, Miller1983rate}
It employs a generalised dividing surface made up of nuclear and electronic projection operators to define reactants and products, and the generalised flux operator.\cite{Lawrence2020NAQI} The location of the generalised dividing surfaces is optimised variationally.

While our NRPI theory is, to our knowledge, the first rigorous nonadiabatic instanton rate theory, it is not the first attempt at calculating a nonadiabatic rate using the steepest-descent approximation.
In fact, the first attempts were presented by Voth and coworkers in the late nineties and they rely on the ImF premise.\cite{Cao1995nonadiabatic, Cao1997nonadiabatic, Schwieters1998diabatic}
Mean-field ring-polymer instanton (MFRPI) theory is the last of a line of instanton rate theories developed by Voth and coworkers and it was revived recently by Ranya and Ananth.\cite{Schwieters1998diabatic, Ranya2020}
MFRPI theory attempted to fix problems identified in its predecessors.
However, as we show in this work, it still only recovers one of the two limits of diabatic coupling, namely the BO one, correctly.
In this study we therefore aim to identify the causes of the issues of MFRPI\@.
We then develop a correction to it, called nonadiabatic ImF (n-ImF) theory, which addresses the main problem in the GR limit.
We thus investigate whether a reliable ImF-based nonadiabatic rate theory which can successfully bridge between the BO and GR limit is possible.
We will start by recapitulating the ImF approach in Sec.~\ref{sec:ImF}, and MFRPI theory in Sec.~\ref{sec:MFRPI}.
In Sec.~\ref{sec:NImF} we introduce our new n-ImF theory and we discuss its application in Sec.~\ref{sec:results}, followed by the conclusions of this paper in Sec.~\ref{sec:conclusions}.
\section{$\text{ImF}$ premise in the BO limit}
\label{sec:ImF}
The derivation of instanton theory from the ImF premise is
rooted in the notion that the ``free energy'' $F$ acquires an imaginary contribution in the barrier region, and that this contribution can be related to the rate of barrier crossing within the semiclassical approximation.\cite{Langer1967ImF, Langer1969ImF, Coleman1977ImF, Callan1977ImF, UsesofInstantons, Affleck1981ImF, Benderskii, Cao1996QTST, RPInst}

For simplicity, we will present the derivations for a one-dimensional system, as the multidimensional extension is easily obtained.
\cite{InstReview}
For a system in the BO approx with a Hamiltonian given by
\begin{equation}
\Hop_\text{BO} = \frac{\op{p}^2}{2m} + V_\text{BO}(\op{x}),
\label{eq:H_BO}
\end{equation}
the partition function can be written in path-integral form as\cite{Feynman}
\begin{align}
    Z = \Tr\left[\eu{-\beta\Hop_\text{BO}}\right] = \lim_{N\rightarrow\infty} \Lambda^{-N} \int \text{d$\bm{x}$ } \eu{-\beta_N U^\text{RP}(\bm{x})},
\end{align}
where $\Lambda = \sqrt{2\pi\beta_N\hbar^2/m}$.
The corresponding ring-polymer potential is
\begin{align}
U_\text{RP}(\bm{x}) &= U_\text{springs}(\bm{x}) + \sum_{i=0}^{N-1} V_\text{BO}(x_i) \label{eq:PES_terms}
\end{align}
where the spring term is defined as
\begin{align}
U_\text{springs}(\bm{x}) &= \sum_{i=0}^{N-1} \frac{m}{2\beta_N^2 \hbar^2} (x_i - x_{i-1})^2.
\label{eq:URP_BO}
\end{align}
We have introduced $\beta_N = \beta/N$ as the effective inverse temperature and $\bm{x} = \{x_1, \dots, x_N\}$ as the $N$ bead coordinates with periodic indexing such that $x_0 = x_N$.
Taking a steepest-descent approximation around a path $\tilde{\bm{x}}_\text{R}$ collapsed at the reactant minimum gives the reactant partition function as
\begin{equation}
Z_\text{R} \simeq \prod_{k=0}^{N-1} \frac{1}{\beta_N\hbar\omega_k} \ \eu{-\beta_N U_\text{RP}(\tilde{\bm{x}}_\text{R})}.
\label{eq:reactionpartitionfunction}
\end{equation}
The frequencies $\omega_k$ are defined as the square roots of the eigenvalues of the mass-weighted ring-polymer Hessian $\grad^2U_\text{RP}(\tilde{\bm{x}}_\text{R})/m$.
In the $N \rightarrow \infty$ limit, the expression can also be evaluated analytically to give the well-known harmonic-oscillator partition function\cite{Kleinert,InstReview}
\begin{equation}
Z_\text{R} = \frac{\eu{-\beta V_\text{R}}} {2\sinh\left(\half\beta\hbar\omega_\text{R}\right)},
\end{equation}
where $V_\text{R}$ is the potential in the reactant well and $\omega_R$ its frequency.
However, the $N$-bead approximation for the reactant partition function $Z_\text{R}$ as shown in Equ.~\eqref{eq:reactionpartitionfunction} is used in the following, as this is in line with the other approximations of ring-polymer instanton theory and thus leads to beneficial convergence properties.\cite{InstReview}
The ImF rate expression depends on the imaginary part of the free energy $F = -\frac1\beta \ln{Z}$ as discussed above, which can be related to the real and imaginary parts of the partition function $Z$ as 
\begin{equation}
k \approx -\frac{2}{\hbar} \Im{F} \approx \frac{2}{\beta \hbar} \frac{\Im {Z}}{\Re{Z}}.
\label{eq:ImF_rate}
\end{equation}
The reactant partition function is identified here with $\Re Z$.
The imaginary part of the partition function of the system in the low-temperature regime is obtained using the saddle point $\tilde{\bm{x}}$, of $U_\text{RP}$, also known as the instanton or optimal tunnelling pathway.
Using the steepest-descent approximation, the imaginary part of the partition function is\cite{InstReview}
\begin{equation}
    \Im{Z} \simeq \frac{N}{2}\sqrt\frac{B_N}{2 \pi \beta_N \hbar^2} \sideset{}{'}\prod_{k=1}^{N-1} \frac{1}{\beta_N\hbar|\omega_k|} \, \mathrm{e}^{-\beta_N U_\text{RP}(\tilde{\bm{x}})},
    \label{eq:ImZ}
\end{equation}
where %
$B_N = \sum_i m (\tilde{x}_i - \tilde{x}_{i+1})^2$ and the prime attached to the product over frequencies indicates the exclusion of the zero-frequency mode (which arises from the permutational degree of freedom of the ring polymer).
The integration over the zero-frequency mode contributes $\sqrt{B_N}$ to the prefactor.
Here, the frequencies $\omega_k$ are defined as the square roots of the eigenvalues of the mass-weighted ring-polymer Hessian $\grad^2U_\text{RP}(\tilde{\bm{x}})/m$.

Although the ImF premise has not been rigorously derived, one can show that the resulting instanton method is in agreement with the semiclassical instanton theory derived from first principles.\cite{Miller1975semiclassical, Althorpe2011ImF, AdiabaticGreens, InstReview}
However, it is clear that Equ.~\eqref{eq:ImZ} only holds for low temperatures where the instanton solution exists and
the expression diverges at the crossover temperature $T_\text{c}$ when the instanton collapses.
For temperatures above the crossover temperature we must therefore adapt the rate formula.
Affleck was the first to tackle this problem and proposed an \emph{ad hoc} extension of the theory.\cite{Affleck1981ImF}
We will focus on this simple approach in this work, although we note that more recently, uniform semiclassical theories have been proposed to bridge the deep-tunnelling and the high-temperature regime.\cite{Pollak2024,Lawrence2024uniform}
\subsection{Affleck's theory for high temperatures}
\label{subsubsec:Affleck}
Affleck's theory extends the ImF-rate expression from the deep-tunnelling regime to the high-temperature limit.\cite{Affleck1981ImF,Cao1996QTST}
The transition from low to high temperatures encompasses a change of mechanism.
Above the crossover temperature, the saddle point corresponds to a ring polymer collapsed at the top of the barrier, $x^\ddagger$. %
In this case, the imaginary part of the partition function can be approximated by
\begin{equation}
    \Im{Z} \simeq \frac{1}{2} \prod_{k=0}^{N-1} |\beta_N\hbar\omega_k|^{-1} \, \mathrm{e}^{-\beta_N U_\text{RP}(\bm{x}^\ddagger)}.
\end{equation}
The sum is now taken over all $N$ modes, since this expression no longer has a zero-frequency mode.

The ring polymer therefore only explores the region of the barrier top and should thus give the well-known rate for the parabolic barrier.\cite{BellBook} %
This is achieved using the following expression whenever one is above the crossover temperature: 
\begin{equation}
    k \approx -\frac{ \beta \omega_\text{b}}{\pi} \Im{F},
    \label{eq:Affleck_rate}
\end{equation}
where the barrier frequency is $\omega_\text{b} =\sqrt{-\nabla^2V_\text{BO}(x^\ddag)/m}$.
Note that the functional form of Equ.~\eqref{eq:ImF_rate} and Equ.~\eqref{eq:Affleck_rate} match at $\beta = \beta_\text{c}$, where inverse crossover temperature is $\beta_\text{c} = 2 \pi/\hbar \omega_\text{b}$.

\section{Mean-field ring-polymer instanton theory}
\label{sec:MFRPI}

MFRPI theory as originally proposed by Schwieters and Voth\cite{Schwieters1998diabatic} was the culmination of a series of developments.\cite{Cao1995nonadiabatic, Cao1997nonadiabatic}
The name \emph{mean-field} ring-polymer instanton theory\cite{Ranya2020} originates from taking the trace over the product of matrix exponentials, where the path integral in the diabatic framework is in the matrix representation.
Note that the theory is typically formulated in the diabatic representation for two electronic states, but it could in principle be reformulated in the adiabatic representation and extended to more states.
The Hamiltonian in the diabatic representation is
\begin{equation}
\Hop = \frac{\hat{p}^2}{2 m} + \mathbf{V}(\hat{x})
\label{eq:dia_ham}
\end{equation}
with the potential energy matrix $\mathbf{V}(x)$ given by
\begin{equation}
\mathbf{V}(x) = \begin{pmatrix}
V_0(x) & \Delta(x) \\
\Delta(x) & V_1(x)
\end{pmatrix}. \label{eq:matrix_exp_pot}
\end{equation}
The RP potential for MFRPI theory is defined in analogy to Equ.~\eqref{eq:PES_terms} as\cite{Alexander2001diabatic}
\begin{equation}
    U_\text{RP}(\bm{x}) = U_\text{springs}(\bm{x}) + U_\text{MF}(\bm{x})
\label{eq:MF_RP_potential}.
\end{equation}
The potential part is given by
\begin{align}
   U_\text{MF}(\bm{x}) &= -\frac{1}{\beta_N} \ln{\big(\text{Tr}\left[\mathbf{M}_1 \mathbf{M}_2 ... \mathbf{M}_N \right]\big)} ,
   \label{eq:MF_PES}
\end{align}
where the trace over electronic coordinates contains the ordered product of matrix exponentials. %
Each bead has an associated matrix exponential given by 
\begin{equation}
\mathbf{M}_i = \eu{-\beta_N \mathbf{V}(x_i)}
\end{equation}
with the diabatic potential matrix $\mathbf{V}(x_i)$ given in Equ.~\eqref{eq:matrix_exp_pot}.
The instanton, $\tilde{\bm{x}}$, is then found by identifying the saddle point of $U_\text{MF}(\bm{x})$.
MFRPI theory is then developed in analogy to the deep-tunnelling rate expression of the ImF method introduced in Sec.~\ref{sec:ImF}. 
The imaginary part of the partition function using the MF potential energy is identical to Equ.~\eqref{eq:PES_terms} except $U_\text{BO}(\bm{x})$ is replaced by $U_\text{MF}(\bm{x})$.\cite{Schwieters1998diabatic, Ranya2020}
In this approach, the RP also has a zero-mode originating from permutational invariance.
It is excluded from the product over frequencies $\omega_k$ and is instead treated by explicit integration leading to the same prefactor as before.
\subsection{MFRPI theory in the BO and GR limit}
\label{sec:MFRPI_limits}
\begin{figure*}
    \centering
    \includegraphics[width=1\textwidth]{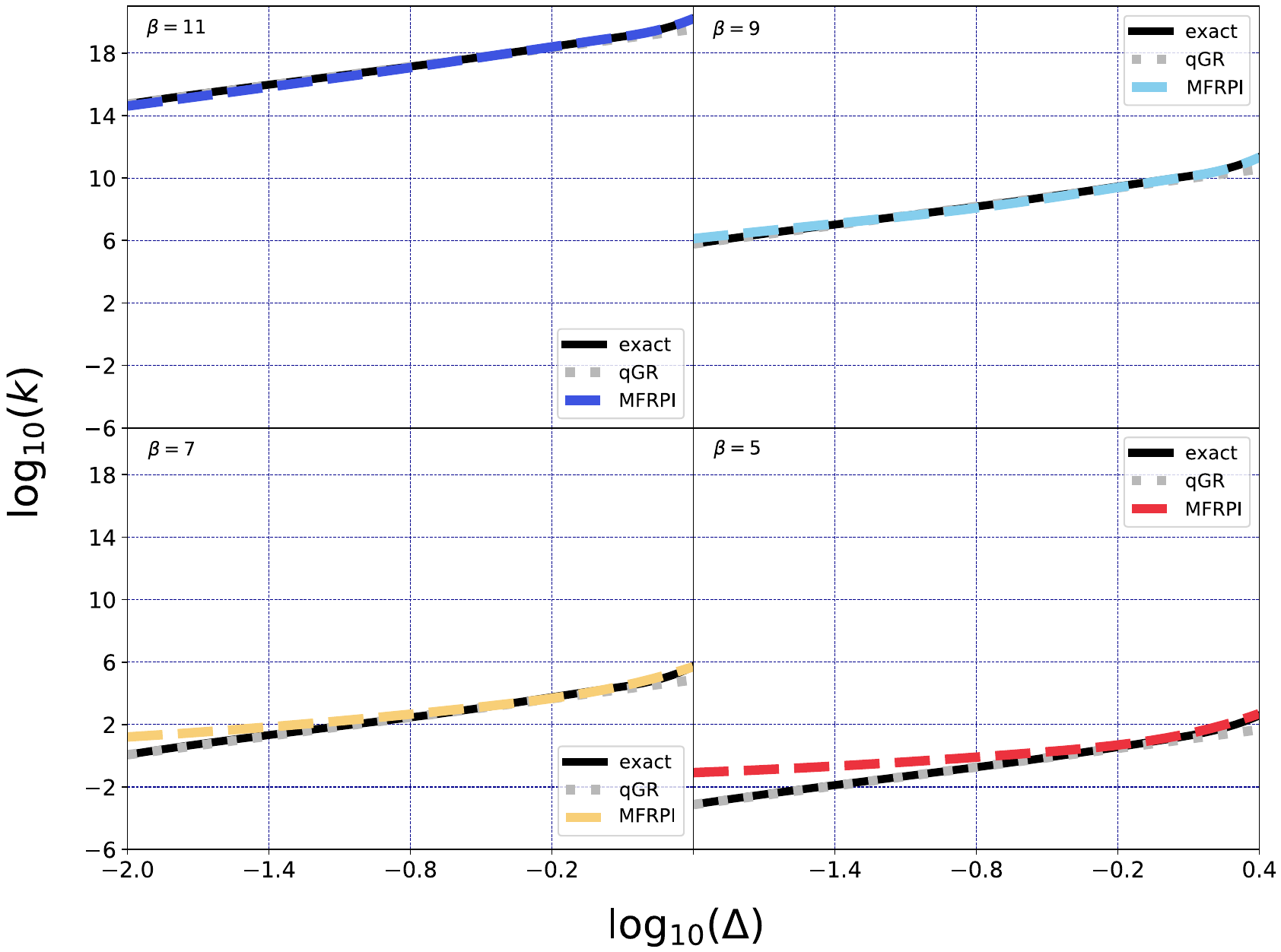}
    \captionsetup{justification=raggedright, singlelinecheck=false}
    \caption{{Asymmetric linear-crossing model with $\kappa_0=1$ and $\kappa_1 = -10$ at different temperatures from low to high. 
    Numerically exact rate (black), numerically-exact quantum GR (dotted grey), MFRPI rate (coloured).
    The error made by MFRPI theory increases with increasing temperature and the $\Delta^2$ behaviour of Fermi's GR is not captured.
    }}
    \label{fig:MFRPI_breakdown_asym_error}
\end{figure*}

A successful nonadiabatic rate theory should recover both the Born--Oppenheimer and the golden-rule limit for strong and weak diabatic coupling, respectively.
In the BO limit, we can show that MFRPI theory rigorously tends to BO instanton theory.
This is achieved by exploiting the diagonal matrix form to reduce the product of matrix exponentials to a simple exponential of a sum of the diagonal terms:
\begin{align}
U_\text{MF}(x) &= \Tr[\mathbf{U}_1\mathbf{U}_1^T \mathbf{M}_1 \mathbf{U}_1\mathbf{U}_1^T \mathbf{U}_2\mathbf{U}_2^T \mathbf{M}_2 \mathbf{U}_2\mathbf{U}_2^T \cdots] \nonumber \\
&= \Tr[\mathbf{U}_1{\mathbf{D}}_1 \mathbf{U}_1^T \mathbf{U}_2 {\mathbf{D}}_2 \mathbf{U}_2^T \cdots]. \nonumber
\end{align}
where ${\mathbf{D}}_i=\mathbf{U}_i^T\mathbf{M}_i\mathbf{U}_i$ is diagonal.
In the case of strong diabatic coupling, one of the on-diagonal terms dominates and the eigenvectors change slowly such that $\mathbf{U}_i^T\mathbf{U}_{i+1} \simeq \mathbf{I}$.
At low temperature, the larger eigenvalue becomes virtually inaccessible, which is equivalent to taking the BO approximation
\begin{align}
U_\text{MF}(\bm{x})
&\simeq \ln{\left(\exp{\left(\sum^N_{i=1} V_\text{BO}(x_i)\right)}\right)} = U_\text{BO}(\bm{x}),
\end{align}
where $V_\text{BO}(x)$ is the lower eigenvalue of $\mathbf{V}(\bm{x})$.
In the limit of strong diabatic coupling, we therefore rigorously recover the RP potential in the BO approximation as given in Equ.~\eqref{eq:URP_BO} and thus also the BO instanton rate prediction. 
In the GR limit ($\Delta \rightarrow 0$), the quantum-mechanical rate is proportional to $\Delta^2$ (assuming constant $\Delta$), where $\Delta$ is the diabatic coupling.
However, we have discovered that MFRPI may completely break down in this limit.
The asymmetric linear-crossing model defined by the diabatic potentials $V_0(x) = \kappa_0 x$ and $V_1(x) = \kappa_1 x$ with parameters $\kappa_0 = 1$ and $\kappa_1 = -10$ provides a simple test case highlighting the issues of MFRPI theory.
The rate coefficients predicted by MFRPI theory are compared to the exact and quantum GR rate in Fig.~\ref{fig:MFRPI_breakdown_asym_error}.
It is clear that the MFRPI prediction in the limit of small $\Delta$ can differ significantly from the exact rate.
This is not just a quantitative failure, but a qualitatively incorrect description of the $\Delta^2$ dependence of the rate in the golden-rule limit.
The problem is encountered sooner at higher temperatures, but exists at all temperatures if one goes to even smaller couplings.
What is particularly concerning is that when computing the MFRPI rate, there is no indication that the breakdown is occurring as the instanton continues to exist, does not collapse and displays a well-defined zero-frequency mode corresponding to the permutational invariance.

The problem is not even limited to asymmetric systems, although the problems are less severe in symmetric cases.
However, some indications of the break-down of MFRPI theory can also be observed for a symmetric one-dimensional linear-crossing model.
Firstly, MFRPI theory exhibits a crossover temperature not only in the adiabatic limit where this is a well-understood phenomenon\cite{InstReview, Affleck1981ImF, Lawrence2024uniform} but also in the GR limit.
At the crossover temperature the MFRP instanton collapses at the top of the barrier and a quantum rate prediction cannot be made beyond this point.
However, from GR instanton theory,\cite{GoldenGreens}
it is known that the instanton should not collapse and indeed predicts small nuclear tunnelling effects even at high temperatures.\cite{Richardson2024}
This effect cannot be captured by a collapsed MFRP instanton.
Additionally, this unphysical crossover temperature depends on the value of $\Delta$, which is clearly inconsistent with the fact that the tunnelling corrections should be independent of $\Delta$ in this limit.
\section{A nonadiabatic $\text{ImF}$ rate theory}
\label{sec:NImF}

We have shown that a na\"ive ``mean-field'' implementation of RP instanton theory does not lead to a reliable rate theory.
We will now show that its failing can be related to contributions from what we refer to as the ``zero-hop'' term which erroneously dominates the rate in the GR limit.
It will be discussed in more detail in the following but it is precisely for this reason that MFRPI theory breaks down in the GR limit as described in Sec.~\ref{sec:MFRPI_limits} (see Appendix~\ref{app:other_attempts} for a discussion of previous attempts leading up to the development of MFRPI theory).
We have already developed a rigorous nonadiabatic instanton rate theory which allows us to obtain an accurate semiclassical rate prediction.\cite{Zarotiadis2025}
Nevertheless, it is interesting to investigate whether a similarly accurate instanton rate theory can be developed using the less rigorous ImF premise.
This would also have the advantage to require only one instanton compared to multiple as required for our rigorous generalised instanton rate theory.
To this end, we thus develop a nonadiabatic ImF (n-ImF) rate theory which inherits ideas from MFRPI theory.
While it is still an \textit{ad hoc} approximation to the exact rate, it appears to give a more reliable rate prediction over a broad range of coupling strengths compared to all previous non-rigorous semiclassical attempts.
The fundamental idea for our correction to the mean-field theory lies in removing the zero-hop contribution such that it cannot falsely dominate the rate.
This can be justified by studying the exact quantum rate expression, which is known to obey an expansion of the form $k = \Delta^2 k_2 + \Delta^4 k_4 + \cdots$.\cite{4thorder}
Each factor of $\Delta$ corresponds to a hop in the ring polymer, implying that there should not be a zero-hop term.
This argument is the justification for the \textit{ad hoc} correction to a mean-field instanton approach presented here.
We therefore propose a new nonadiabatic ImF rate (called n-ImF) based on the ring-polymer potential
\begin{align}
    U^\text{n-ImF}(\bm{x}) &= -\frac{1}{\beta_N} \ln\Bigg(\Tr{\left[\prod_{i=1}^N \mathbf{M}_i \right]} %
    -
    \Tr{\left[\prod_{i=1}^N \mathbf{M}^{(0)}_i\right]}\Bigg). 
   \label{eq:NIMF_PES}
\end{align}
Here, we have introduced the matrices
\begin{equation}
    \mathbf{M}_i^{(0)} =
    \begin{pmatrix} 
        \mathrm{e}^{-\beta V_0(x_i)} & 0 \\
        0 & \mathrm{e}^{-\beta V_1(x_i)} \\
    \end{pmatrix}.
\end{equation}
The second term in Equ.~\eqref{eq:NIMF_PES} corresponds to the contribution which encodes no change or ``hop'' between diabatic states.
In this way, the contributions from this unphysical zero-hop term are eliminated.
\begin{figure}[H]
    \centering \includegraphics[width=1\columnwidth]{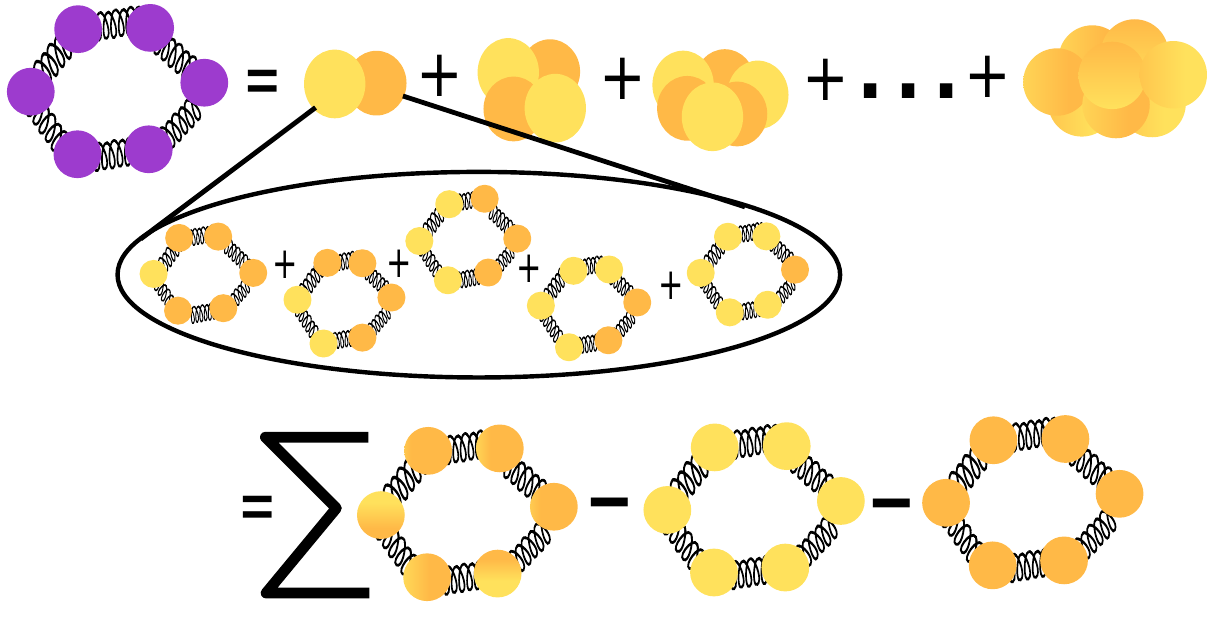}
\captionsetup{justification=raggedright, singlelinecheck=false}
    \caption{{
Cartoon of the contributions to the ``n-ImF'' instanton.
The zero-hop terms are excluded by definition but all higher-order terms are included in a mean-field way.
The n-ImF ring polymer is shown in purple and the different contributions are coloured according to the electronic states $\ket{0}$ (yellow) and $\ket{1}$ (orange).
The zoom on the second-order contribution highlights that all possible ways of describing two hops are accounted for.
}
}
\label{fig:nImF_concept}
\end{figure}

We argued in Sec.~\ref{sec:MFRPI_limits} that MFRPI recovers the BO instanton theory in the appropriate limit.
The modified theory also follows this good behaviour.
This is because the zero-hop term has a negligible contribution in this case, due to the fact including hops can significantly lower the action when the coupling is strong.
These theoretical considerations are also supported by numerical investigation of the MFRPI and the n-ImF rate in the BO limit for a symmetric and an asymmetric system where very good agreement was observed.
Nevertheless, although the n-ImF theory gives reasonable results in the golden-rule limit, it does not formally reduce to the GR instanton theory, which is rigorously derived as an asymptotic semiclassical approximation to the quantum golden-rule rate.
In particular, the n-ImF theory has a crossover temperature at which the instanton collapses for all diabatic couplings.
It thus fails to smoothly describe the onset of tunnelling in the GR limit.
However, this behaviour is no worse than the standard instanton theory in the BO limit.
In this work, we therefore mostly focus on the deep-tunnelling regime, although an attempt is made in Sec.~\ref{subsec:highTnimf} to extend the n-ImF approach to the high-temperature regime.
As for all other instanton theories, it is easy to extend the theory to multidimensional systems.
However, it is advantageous from a computational standpoint to first rewrite the RP potential of n-ImF theory as
\begin{equation}
    U^\text{n-ImF}(\bm{x}) = -\frac{1}{\beta_N}\left(\sum_{i=1}^N \Tr{\left[\mathbf{M}_1^{(0)}\cdots\mathbf{M}_{i-1}^{(0)}\widetilde{\mathbf{M}}_i\cdots\mathbf{M}_N\right]} \right),\label{eq:NIMF_PES_stable}
\end{equation}
where a new matrix $\widetilde{\mathbf{M}}_i$ is defined as
\begin{equation}
    \widetilde{\mathbf{M}}_i = \mathbf{M}_i - \mathbf{M}^{(0)}_i. 
\end{equation}
This is formally equivalent to Equ.~\eqref{eq:NIMF_PES} but it avoids numerical problems when subtracting two traces of large but similar magnitude.
Instead, it is a sum of terms where one forces one hop explicitly and the second hop is forced implicitly by the trace condition.
The second hop is thus free to occur at any other bead and it retains the symmetry of the ring polymer.
Derivatives in terms of nuclear coordinates can be implemented according to Bell's algorithm.\cite{TimMasters, Menzeleev2014kinetic}
\subsection{Connection to the high-temperature limit}
\label{subsec:highTnimf}
Here, we aim to generalise Affleck's idea (see Sec.~\ref{subsubsec:Affleck}) to extend our deep-tunnelling n-ImF theory to obtain a high-temperature rate expression.
The rate expression in Equ.~\eqref{eq:Affleck_rate} can be rewritten equivalently as
\begin{equation}
    k \approx \frac2\hbar \frac{\beta}{\beta_\mathrm{c}} \text{Im} F. \label{eq:ad_cross_T_rate} 
\end{equation}
It can be further generalised to 
\begin{equation}
    k \approx \frac2\hbar \left(\frac{\beta}{\beta_\mathrm{c}}\right) \left(\frac{\beta_\mathrm{c} \hbar \omega_0}{2\pi}\right)^\eta \Im{F},
\end{equation}
where $\eta=1$ and $\eta=0$ can be identified directly as Eqs.~\eqref{eq:Affleck_rate} and~\eqref{eq:ad_cross_T_rate}, respectively.
The imaginary frequency of the RP is denoted as $\omega_0$.
In fact, in the adiabatic limit it holds that $\omega_0 = \omega_\mathrm{b}$ with $\omega_\mathrm{b}$ as the barrier frequency and $\beta_\mathrm{c} =  2 \pi/\hbar \omega_\mathrm{b}$.
In this limit, the rate evaluation is thus independent of the parameter $\eta$.
The formula does however depend on $\eta$, in general.
At sufficiently high temperatures, the instanton only explores a small part of the diabatic potentials, such that it is reasonable to employ a simple approximation. 
To study the high-temperature limit, we therefore consider such a linear-crossing model with potentials $V_n(x) = V^\ddag + \kappa_n x$, where ${n=\{0,1\}}$ identifies the respective diabatic state and $V^{\ddag}$ describes the barrier height.
Given that the rate evaluation is independent of $\eta$ in the BO limit, the aim is to find $\eta$ such that the classical n-ImF rate expression best agrees with the classical golden-rule one in the GR limit.
Analysing the classical GR rate expression for a linear crossing model and assuming a collapsed RP we propose setting $\eta=-2$, which leads to
\begin{equation}
k_\text{cl}^\text{n-ImF} = \frac{2}{\hbar} \frac{\beta}{\beta_\mathrm{c}} \left(\beta_\mathrm{c} \frac{\hbar |\omega_0|}{2\pi}\right)^{-2} \text{Im} F,
\label{eq:nimf_highT}
\end{equation}
using $\omega_0 = \omega_\mathrm{b}$ and $\beta/\beta_\mathrm{c} = 1$ at crossover (see Appendix~\ref{app:highT_nimf} for details).
Our attempt at a high-temperature rate expression relies on a number of assumptions, most notably the collapse of the RP at the crossing point $x=0$, and we will show the resulting limitations in the following.
Nevertheless, it is important to stress that its development is completely separate from that of the deep-tunnelling n-ImF rate theory.
\section{Application of n-ImF theory}
\label{sec:results}
The n-ImF theory can be understood as an improvement upon MFRPI theory.
Here, we apply it to a number of systems from the deep-tunnelling to the classical limit and from the weak- to the strong-coupling regime in both one- to multidimensional systems.
While it does not come without its own weaknesses in certain limits, it always outperforms MFRPI theory by at least capturing the $\Delta^2$-dependence of Fermi's GR. 

\subsection{One-dimensional symmetric model}
\label{subsec:symmetric_lincross}
\begin{figure*}
\centering
\includegraphics[width=1\linewidth]{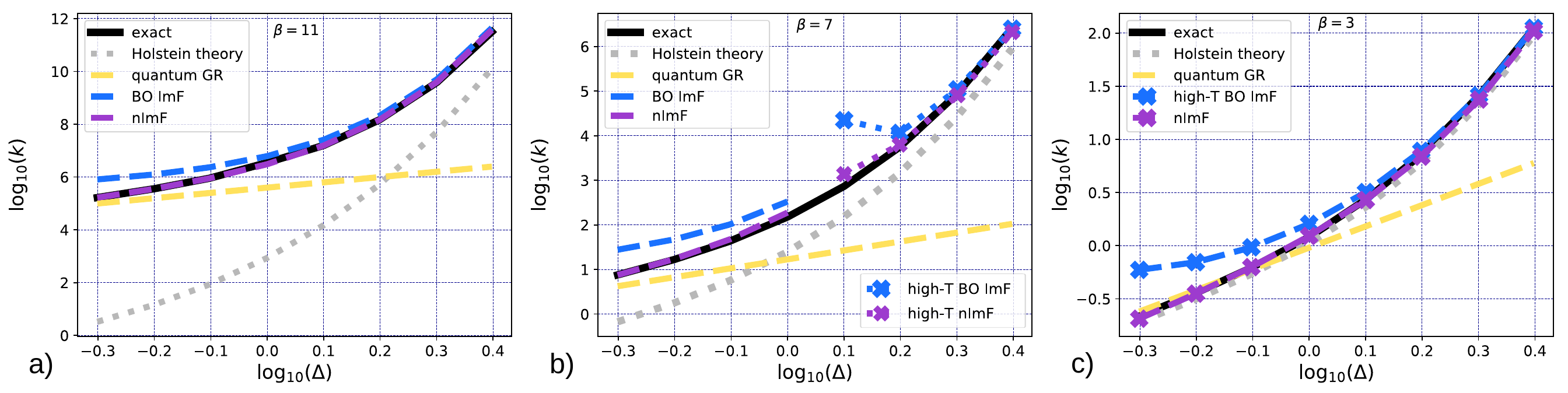}
\captionsetup{justification=raggedright, singlelinecheck=false}
\caption{{Log-log plot of rate coefficients $k$ vs. diabatic coupling $\Delta$ are shown for the quantum-mechanically exact rate (black), Holstein rate (dotted grey), GR rate (yellow), the BO instanton rate (blue) and the n-ImF rate (purple).
The system under study is a symmetric linear crossing model with slopes $\kappa = \pm 1$.
(a) $\beta = 11$. A low-temperature system with the Holstein rate far away from the exact rate. The n-ImF rate successfully predicts the exact rate.
(b) $\beta = 7$. A system in the crossover temperature region, where the crossover between low and high-temperature theories can be observed both for the BO instanton and the n-ImF theory.
The change from low to high temperature formulation of the rate theory leads to a discontinuity at the crossover temperature.
(c) $\beta = 3$. A high-temperature system with Holstein rate close to the exact rate. The high-temperature n-ImF rate predicts the rate well.
}
\label{fig:symlincross}
}
\end{figure*}

We start with two linear-crossing potentials with $\kappa_0=-\kappa_1=1$ coupled by the diabatic coupling $\Delta$.
We first investigated the rate expression for a range of coupling strengths in the deep-tunnelling regime at the inverse temperature $\beta = 11$ with reduced units $m = \hbar = 1$.
The exact rate was obtained by evaluation of the transmission probability via the Greens functions on a grid.
We compare all of our rate predictions to the classical Holstein rate theory\cite{Holstein1959b, Nitzan} (see Appendix~\ref{app:Holstein}) which highlights the extent to which nuclear quantum effects can be observed in a given system.
Fig.~\ref{fig:symlincross}(a) shows that the n-ImF theory predicts the rate constants well for the full range of coupling strengths going from the BO limit with strong coupling to the GR limit with weak coupling, and the large deviation of the classical Holstein rate prediction underlines the deep-tunnelling nature of the systems under study.
The switch from the low-temperature theory [Equ.~\eqref{eq:ImF_rate}] to the high-temperature theory [Equ.~\eqref{eq:nimf_highT}] is indicated by the symbols used to mark the data points in Fig.~\ref{fig:symlincross}(c).
Close to the crossover, the rates diverge from the exact rate, similarly to the standard BO instanton theory.
Finally, the high-temperature limit is shown in Fig.~\ref{fig:symlincross}(b), and all n-ImF rates are obtained from the high-temperature rate expression.
The n-ImF rate predictions match well with the exact rate also in this regime from one limit in diabatic coupling to the other.
In addition, as tunnelling is quenched in this case, the n-ImF theory is in good agreement with the classical Holstein approach.

Overall, we have thus shown that the n-ImF rate theory performs well for a symmetric one-dimensional model in both the low- and high-temperature regimes.

\subsection{One-dimensional asymmetric model}
\label{subsec:nImF_deept}
\begin{figure}
    \centering
    \includegraphics[width=.8\columnwidth]{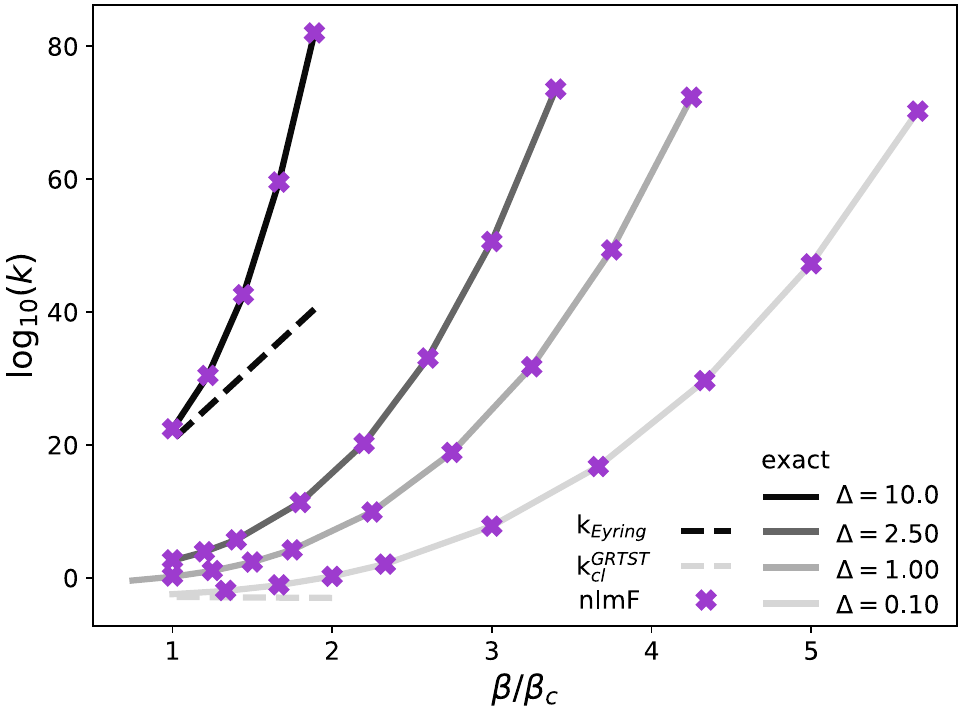}
    \captionsetup{justification=raggedright, singlelinecheck=false}
    \caption{{Rate constants as a function of the inverse temperature $\beta$ for the asymmetric linear-crossing model with diabatic couplings from $\Delta = 10$ (BO limit) to $\Delta = 0.1$ (GR limit).
    The inverse temperature is scaled by $\beta_\mathrm{c}$ of the n-ImF theory for each given $\Delta$.
    Solid lines give the quantum-mechanically exact rate and the purple crosses depict the low-temperature n-ImF rate prediction which shows excellent agreement.
    The dashed lines show the Eyring TST and the classical GR rate predictions for coupling strengths $\Delta = 10$ and $\Delta = 0.1$, respectively and they indicate the systems are all in the deep-tunnelling regime for $\beta > \beta_\mathrm{c}$.
    }}
    \label{fig:asym_lowT}
\end{figure}

Many previously proposed rate theories work well for the symmetric case but fail for asymmetric models.
In order to demonstrate that this is not the case for the n-ImF theory, we thus investigated an asymmetric linear-crossing model with slopes $\kappa_0 = 1$ and $\kappa_1 = -10$.
In the low-temperature limit even for extreme asymmetry the n-ImF theory recovers the exact rates well for all coupling strengths from the GR to the BO limit (see Fig.~\ref{fig:asym_lowT}).
Indeed, the low-temperature rate predictions are successful all the way to the classical limit which can be seen in Fig.~\ref{fig:asym_lowT} by comparison to the dashed classical rate predictions from Eyring theory in the BO limit and the exact classical GR rate as given in Equ.~\eqref{eq:general_1D_highT_GRrate} in the GR limit.
\subsection{Multidimensional symmetric model}
\label{subsec:symmetric_spin_boson_nimf}
The spin--boson model, which we will express in the system--bath form, is the simplest model of a multidimensional nonadiabatic chemical reaction.
One assumes that the chemical reaction takes place along a reaction coordinate $Q$ and the surrounding environment is well described by the $(f-1)$-dimensional bath with nuclear coordinates $q$.
The diabatic potentials are defined as
\begin{equation}
    V_n(Q, q) = \frac{1}{2}\Omega^2\left(Q + (-1)^n \sqrt{\frac{\Lambda}{2\Omega}}\right)^2 + V_\text{sb}(Q, q),
\end{equation}
where $n \in \{0, 1\}$ and the potential $V_\text{sb}(Q, q)$ is given by
\begin{equation}
    V_\text{sb}(Q, q) = \sum_{j=1}^{f-1} \frac{1}{2} \omega_j^2 \left(q_j - \frac{c_j Q}{\omega_j}\right)^2   ,
\end{equation}
which contains the couplings constants $c_j$ and frequencies $\omega_j$.
We choose the Brownian oscillator spectral density to characterise the bath, which is defined as
\begin{equation}
    J_\sigma(\omega) = \frac{\Lambda}{2}\frac{\gamma\Omega^2\omega}{\left(\omega^2-\Omega^2\right) + \gamma^2\omega^2}.
\end{equation}
The parameter $\Omega$ is the frequency along the reaction coordinate $Q$ and the parameter $\Lambda$ can be identified as the Marcus reorganisation energy.
Solvent friction is introduced into the system via the coefficient $\gamma$.
The parameters studied here were taken from Ref.~\onlinecite{Lawrence2019ET}.
The spin--boson model allows us to study the effect of solvent friction in addition to the different regimes of diabatic coupling from the GR to the BO limit.
The regime of solvent friction is determined by comparison of the solvent coefficient $\gamma$ and the frequency along the reaction coordinate $\Omega$.
A system is in the underdamped low friction regime if $\gamma < 2 \Omega$.
In contrast, it is in the overdamped high friction regime if $\gamma > 2 \Omega$.
The characterisation of the diabatic coupling strength of a given system is made by comparison to previous rate calculations which rely on the approximations of either limit.
In particular, ring-polymer molecular dynamics (RPMD) is applicable in the BO limit.\cite{RPMDrate, RPMDrefinedRate, RPInst}
The ratio between RPMD and Eyring TST\cite{Eyring1935review, Eyring1938rate} gives an indication of the importance of tunnelling effects.
In the opposite GR limit, Wolynes theory is applicable.\cite{Wolynes1987nonadiabatic}
In the following, these rate theories are therefore employed as a reference to determine the region of diabatic coupling of a given system.
Additionally numerically-exact hierarchical equations of motion (HEOM) are used to obtain exact reference results.
The results for HEOM, RPMD and Wolynes theory are taken from Ref.~\onlinecite{Lawrence2019ET}.

In order to span the full range from weak to strong solvent friction and also from weak to strong diabatic coupling, a number of different parameter regimes are investigated.
We already discussed in Sec.~\ref{sec:ImF} that
all instanton theories derived from the ``ImF'' premise exhibit a crossover temperature at which the RP collapses and we indicate an above-crossover rate prediction by a ``$\ast$''. 
The collapse of the RP marks the crossover between quantum and classical regime for a given rate theory.
Due to the change of dominant mechanism, the rate prediction close to crossover may be inaccurate.
We present the crossover temperatures of both the BO-instanton and the n-ImF theory in the second section of Tables~\ref{tab:lowfriction_lowT}--\ref{tab:highfriction_highOmega}.
In the BO limit, the inverse crossover temperature $\beta_\mathrm{c}^\text{n-ImF}$ always converges to the BO one as is expected since the high-temperature n-ImF theory recovers BO-instanton theory in this limit (see Subsec.~\ref{subsec:highTnimf}).
For a system with parameters $\beta = 1$, $\Lambda = 60$, $\Omega = 4 \hbar$, $\hbar=1$ and $\gamma = \Omega$, the resulting rates are given in Table~\ref{tab:lowfriction_lowT}.
This system is in the small solvent-friction limit and it highlights the crossover from the low- to high-temperature regime.
The n-ImF rate is in good agreement with the numerically-exact HEOM\cite{Lawrence2019ET} results.
In the BO limit, the comparison of the RPMD rate prediction with the BO instanton rate prediction serves as an estimate of the error of the instanton approximation.
Regarding the crossover temperatures, the BO and \mbox{n-ImF} theories differ strongly in the GR limit since one is based strictly on the BO approximation which breaks down in this regime.
We can compare the crossover temperature of the n-ImF theory in this regime to the crossover temperature of a mean-field version of the GR expression.
This crossover temperature is $\beta_\mathrm{c}^\text{GR, n-ImF} = 0.74$ and therefore smaller than the crossover temperature of the smallest diabatic coupling $\Delta = 0.1$ shown in Table~\ref{tab:lowfriction_lowT}.
This suggests the system is in an intermediate regime with regard to the diabatic coupling strength.
It should be pointed out that the agreement of the interpolation formula (IF) bridging between RPMD and Wolynes theory proposed by Lawrence et al.\cite{Lawrence2019ET} with their numerically-exact HEOM results is high.
There are however two fundamental advantages of the n-ImF theory over the interpolation formula approach. 
Firstly, n-ImF theory is derived from a single, unified origin therefore going beyond the simple combination of scalar rate coefficients.
Secondly, in addition to the rate, it provides an optimal tunnelling pathway which describes the reaction mechanism.
For a more extensive description of the IF method see Appendix~\ref{app:IF}.
\begin{table*}[!htbp]
\captionsetup{justification=raggedright, singlelinecheck=false}
\caption{{Rate predictions given as $\log_{10}(k)$ and crossover temperatures $\beta_\mathrm{c}$ from different rate theories for a system with parameters: $\beta = 1, \Lambda = 60, \Omega = 4 \hbar$ and $\gamma = \Omega$.
All rates marked with a ``$^\dag$'' are taken from Lawrence et al.~\cite{Lawrence2019ET}
All rates are obtained with $f=14$ degrees of freedom and a $N=120$ number of beads besides the Eyring rate which was already converged at $f=5$ and $N=50$ and the high-temperature ``ImF'' rate which was converged at $f=9$ and $N=50$.
The crossover temperatures were obtained for $f=9$ and $N=50$.
Above crossover rates are highlighted with a ``$\ast$''.
}}
\label{tab:lowfriction_lowT}
\begin{tabular}{l|lllllll}
$\Delta$ & 0.10 & 1.00 & 2.51 & 3.16 & 5.00 & 6.31 & 10.0 \\
\hline
$\log_{10}{(k_\text{n-ImF})}$ & -8.05 & -6.02 & -5.06  & -4.74 & -4.06* & -3.79* & -2.86* \\
$\log_{10}{(k_\text{HEOM})}^\dag$ & -8.00 & -5.98  & -5.10  & -4.84 & -4.23 & -3.85 & -2.89  \\
$\log_{10}{(k_\text{IF})}^\dag$                         & -8.00 & -5.97  & -5.05  & -4.79 & -4.18 & -3.81 & -2.87  \\
$\log_{10}{(k_\text{Wolynes})}^\dag$ & -8.00 & -6.00  & -5.20  & -5.00 & -4.60 & -4.40 & -4.00  \\
$\log_{10}{(k_\text{RPMD})}^\dag$                       & -4.78 & -4.72  & -4.49  & -4.37 & -3.96 & -3.66 & -2.80  \\
$\log_{10}{(k_\text{BO-ImF})}$ & --   & -4.60  & -4.34  & -4.18 & -3.63 & -2.60* & -2.59*  \\
$\log_{10}{(k_\text{Eyring})}$ & -5.77 & -5.70 & -5.22 & -5.00 & -4.40 & -3.98 %
& -2.94 \\ %
\hline
$\beta_\mathrm{c}^\text{BO-ImF}$                                & 0.093 & 0.318 & 0.549 & 0.636 & 0.873 & 1.042 & 1.565 \\
$\beta_\mathrm{c}^\text{n-ImF}$                         & 0.920 & 0.926 & 0.958 & 0.982 & 1.083 & 1.188 & 1.625
\end{tabular}
\end{table*}
Rate theories and their predictions are significantly impacted by the introduction of strong solvent friction.
The strong-friction limit is challenging for a number of theories.\cite{Lawrence2019ET, Trenins2020} 
The rate coefficients for a number of rate theories are shown in Table~\ref{tab:highfriction_lowOmega} for $\Omega=0.5\hbar$ and in Table~\ref{tab:highfriction_highOmega} for $\Omega=4\hbar$.
From the RPMD and Wolynes rates, it is clear that the full range of diabatic coupling is covered.
However, all rates are in the high-temperature limit with respect to the ``ImF'' instanton theories meaning all rates are obtained from collapsed RPs.
For the system with $\Omega=0.5\hbar$, this is further supported by the HEOM and the Marcus theory\cite{Marcus1956ET} rates matching in the GR limit.
Secondly, the Eyring and high-temperature BO-ImF rates are close to the exact result for the BO systems ($\Delta \gtrsim 3.16$) in both cases.
For both frequencies, the interpolation between Wolynes and RPMD rate coefficients performs well.\cite{Lawrence2019ET}
The high-temperature extension of the n-ImF theory can approximate the numerically-exact HEOM rates well only in the BO limit where it is rigorously shown to tend to the high-temperature BO-ImF extension (see Secs.~\ref{subsec:highTnimf} and~\ref{subsubsec:Affleck}).
In conclusion, the n-ImF theory performs comparably well for systems in the low-friction, deep-tunnelling regime across all diabatic couplings as it already did for the one-dimensional model studied in Sec.~\ref{subsec:symmetric_lincross}.
Our attempt at a high-temperature extension of the n-ImF theory captures the rates in the high-friction limit in the case of strong coupling.
Beyond the BO limit however, it fails to reproduce the rates when high solvent friction is studied.
It is important to note that any relevant mispredictions by n-ImF theory have been made by the high-temperature rate expression only.
It is based on a collapsed RP and its derivation is strictly separate from the deep-tunnelling ImF rate theory for which all predictions show at most a small error.
The (deep-tunnelling) n-ImF theory can thus be said to perform well, while a high-temperature rate theory requires further study and improvement.
\begin{table*}[!htbp]
\captionsetup{justification=raggedright, singlelinecheck=false}
\caption{{Rate predictions given as $\log_{10}(k)$ and crossover temperatures $\beta_\mathrm{c}$ from different rate theories for a system with parameters: $\beta = 1, \Lambda = 60, \Omega = 0.5 \hbar$ and $\gamma = 32 \Omega$.
All rates marked with a ``$^\dag$'' are taken from Lawrence et al.~\cite{Lawrence2019ET}
All rate coefficients are converged with respect to the number of beads at a given number of degrees of freedom.
The high-temperature ``ImF'' rate was converged at $f=11$ and $N=100$.
The GR-n-ImF crossover temperature is $\beta_\mathrm{c}^\text{GR, n-ImF} = 2.96$ and therefore much smaller than the smallest n-ImF crossover temperature.
Above crossover rates are highlighted with a ``$\ast$''.
The $\diamond$ indicates inverse crossover temperatures which were not explicitly computed since they are expected to converge to the BO crossover temperatures. 
}}
\label{tab:highfriction_lowOmega}
\begin{tabular}{l|llllllll}
$\Delta$ & 0.10 & 0.40 & 1.00 & 2.51 & 3.16 & 5.00 & 6.31 & 10.0 \\
\hline
$\log_{10}{(k_\text{n-ImF})}$           & -8.45*& -7.37* & -7.07* & -7.19* & -7.11* & -6.71* & -6.36* & -5.37* \\
$\log_{10}{(k_\text{HEOM})}^\dag$            & -9.19 & -8.36 & -8.10 & -7.62 & -7.41 & -6.82 & -6.42 & -5.41   \\ 
$\log_{10}{(k_\text{IF})}^\dag$              & -9.20 & -8.38 & -8.11 & -7.62 & -7.41 & -6.83 & -6.43 & -5.39 \\
$\log_{10}{(k_\text{Wolynes})}^\dag$         & -9.14 & -7.94 & -7.14 & -6.34 & -7.41 & -5.74 & -5.54 & -5.14 \\
$\log_{10}{(k_\text{RPMD})}^\dag$            & -8.25 & -8.21 & -8.07 & -7.62 & -7.41 & -6.83 & -6.43 & -5.39 \\
$\log_{10}{(k_\text{BO-ImF})}$              & --  & -7.63* & -7.85* & -7.52* & -7.32* & -6.77* & -6.37* & -5.37* \\
$\log_{10}{(k_\text{MT})}$              & -9.15& -7.95 & -7.15 & -6.36 & -6.16  & -5.76 & -5.55 & -5.15 \\
$\log_{10}{(k_\text{Eyring})}$ & -7.32 & -7.89 & -7.81 & -7.50 & -7.31 & -6.76 & -6.37 & -5.33 \\
\hline
$\beta_\mathrm{c}^\text{BO-ImF}$                     & -- & 4.00 &14.15 & 37.00 & 47.64 & 80.71 & 107.40 & 201.36 \\
$\beta_\mathrm{c}^\text{n-ImF}$              & 8.39 & 9.25 &14.48 & 36.99 & $\diamond$ & $\diamond$ & $\diamond$ & $\diamond$
\end{tabular}
\end{table*}

\begin{table*}[!htbp]
\captionsetup{justification=raggedright, singlelinecheck=false}
\caption{{Rate predictions given as $\log_{10}(k)$ and crossover temperatures $\beta_\mathrm{c}$ from different rate theories for a system with parameters: $\beta = 1, \Lambda = 60, \Omega = 4 \hbar$ and $\gamma = 32 \Omega$.
All rates marked with a ``$^\dag$'' are taken from Lawrence et al.~\cite{Lawrence2019ET}
All rate coefficients are converged with respect to the number of beads at a given number of degrees of freedom.
The GR-n-ImF crossover temperature is $\beta_\mathrm{c}^\text{GR, n-ImF} = 0.74$ as discussed before.
If the diabatic coupling is weak the BO instanton theory runs into convergence issues and therefore is not given which is indicated by ``--''.
Above crossover rates are highlighted with a ``$\ast$''.
The $\diamond$ indicates that the crossover temperatures are expected to converge to the BO-ImF equivalents and are therefore not calculated. 
}}
\label{tab:highfriction_highOmega}
\begin{tabular}{l|lllllll} 
$\Delta$ & 0.10 & 1.00 & 2.51 & 3.16 & 5.00 & 6.31 & 10.0 \\
\hline
$\log_{10}{(k_\text{n-ImF})}$           & -8.86* & -6.89* & -6.32* & -6.20* & -5.77* & -5.42* & -4.43* \\
$\log_{10}{(k_\text{HEOM})}$            & -9.05 & -7.29 & -6.69 & -6.46 & -5.86 & -5.47 & -4.46  \\
$\log_{10}{(k_\text{IF})}$              & -9.05 & -7.22 & -6.60 & -6.39 & -5.82 & -5.43 & -4.41 \\
$\log_{10}{(k_\text{Wolynes})}$         & -9.05 & -7.05 & -6.25 & -6.05 &-5.65 & -5.45  & -5.05 \\
$\log_{10}{(k_\text{RPMD})}$            & -7.04 & -6.91 & -6.54 & -6.35 & -5.80 & -5.42 & -4.41 \\
$\log_{10}{(k_\text{BO-ImF})}$              & -- & -5.99*  & -6.27*  & -6.15*  & -5.69*  & -5.34* & -4.38*   \\
$\log_{10}{(k_\text{Eyring})}$          & -2.36 & -6.25 & -6.32 & -6.17 & -5.70 & -5.34 & -4.38 \\
\hline
$\beta_\mathrm{c}^\text{BO-ImF}$                     & --  & 1.77   & 4.62   & 5.95   & 10.10  & 13.44  & 24.61  \\
$\beta_\mathrm{c}^\text{n-ImF}$              & 3.0   & 3.24   & 4.97   & 6.15   & 10.16  &  $\diamond$    & $\diamond$     \\
\end{tabular}
\end{table*}
\subsection{Understanding the limitations of n-ImF theory}
\label{subsec:nImF_asym_lin_cross}
\begin{figure*}[!htbp]
    \centering
    \includegraphics[width=1\linewidth]{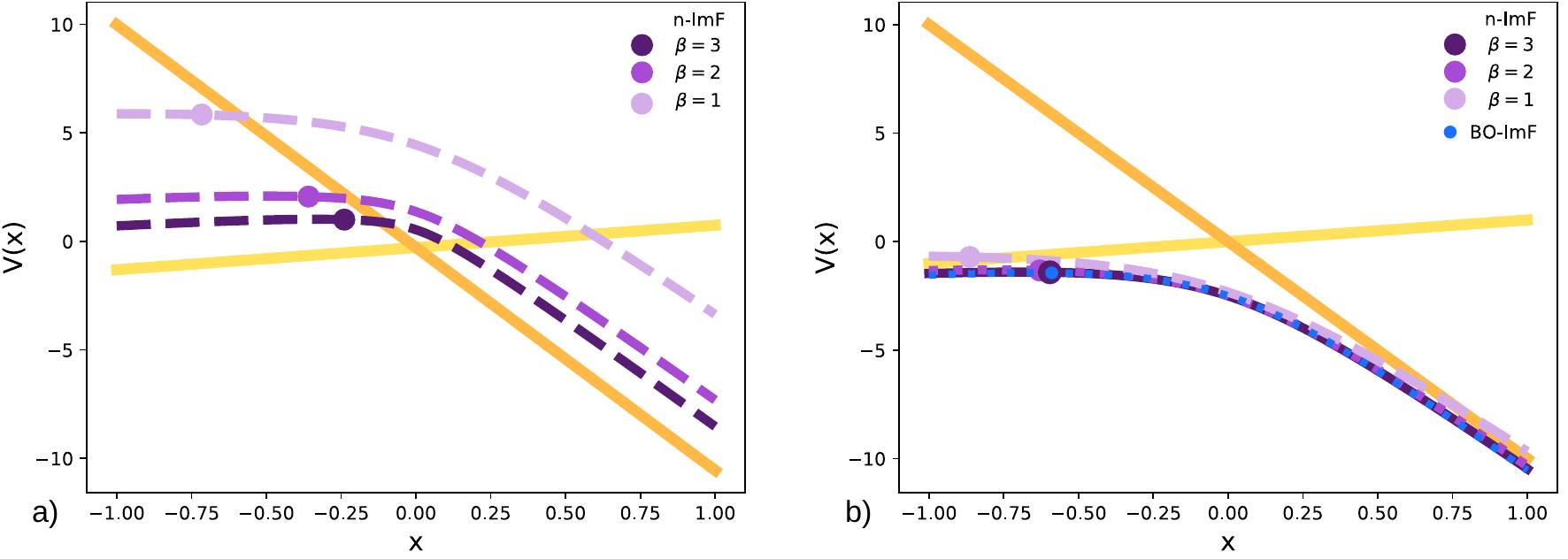}
    \captionsetup{justification=raggedright, singlelinecheck=false}
    \caption{{Asymmetric linear-crossing model with $\kappa_0 = 1 $ and $\kappa_1 = -10$. Plot of the diabatic states $V_n(x)$ along the nuclear coordinate $x$ in yellow and orange and the effective MFRPI surfaces at different temperatures in shades of purple.
    The BO surface is shown in blue in the strong coupling case in Panel~(b).
    (a) $\Delta = 0.1$, system in the GR limit. Collapsing points of the n-ImF rates are strongly affected by temperature.
    (b) $\Delta = 2.5$, system in the BO limit. Collapsing points of the n-ImF theory in purples and the BO instanton in blue. The deviation between the BO and the n-ImF collapsing points is small.
    }}
    \label{fig:asym_runaway}
\end{figure*}

In order to understand why our attempt at an extension of the n-ImF theory to the high-temperature regime is unsuccessful in the GR limit we return to the one-dimensional asymmetric linear-crossing model.
For this model, introduced in Sec.~\ref{subsec:nImF_deept}, our attempt at a nonadiabatic high-temperature extension cannot capture the correct behaviour of the classical path in the GR limit.
In fact, the location of the collapsed RP changes significantly relative to the inverse temperature $\beta$ (see Fig.~\ref{fig:asym_runaway}a)).
As the temperature increases the point of collapse of the RP moves away from the crossing point, the expected transition state.
As a consequence, the rate is measured at an ill-chosen point and the rate prediction is inaccurate.
This can also be understood in the context of GR instanton theory where the two changes or hops between electronic states are optimised by steepest descent and only a single, optimal configuration contributes to the rate coefficient.
In the n-ImF theory instead we average over all possible locations of the two hops as is illustrated in the cartoon of Fig.~\ref{fig:nImF_concept}.
The inclusion of all these contributions prevents the high-temperature extension presented here from capturing the rate correctly.
In fact, a similar failing can also be observed in the BO limit.
Here also, the n-ImF high-temperature extension is $\beta$-dependent and thus unphysical since we know from BO instanton theory that the point of collapse is $\beta$-independent in the BO limit (see Fig.~\ref{fig:asym_runaway}b)).
However, in the BO limit the n-ImF rate is not dominated by the changes of electronic state.
The difference between the location of collapse from the high-temperature n-ImF and BO theory is hence small which makes the rate prediction more accurate in the BO limit compared to the GR limit.
Already for the symmetric multidimensional spin--boson model in the high-friction regime, we showed that the rate predictions from the high-temperature extension deviate from the exact result in the GR limit.
This can be understood by relating the high-friction limit to the high-temperature limit in the sense that in both limits nuclear tunnelling effects become subdominant compared to classical thermodynamic fluctuations.

\section{Conclusions}
\label{sec:conclusions}
We have given an overview of existing efforts towards a nonadiabatic rate theory using the ImF premise, especially MFRPI theory.\cite{Schwieters1998diabatic, Ranya2020}
We showed that it rigorously recovers the BO instanton theory in the appropriate limit.
In the opposite limit, however, MFRPI theory breaks down in an uncontrollable fashion by exhibiting premature collapse of the RP and erroneous rate predictions.
This break-down leads to inaccurate predictions of the reaction rate by orders of magnitude.
As an alternative, we have recently developed the NRPI rate theory which enables bridging between the BO and the GR limit in a rigorous manner.\cite{Zarotiadis2025}
As mentioned earlier, our NRPI theory has been developed in the flux-correlation framework which has proven to be a successful starting point.
Nevertheless, in the BO limit, one can show that the ImF instanton expression is equivalent to that derived from the flux correlation function.\cite{Althorpe2011ImF, AdiabaticGreens, InstReview}
We thus aimed to investigate whether a superseding rate theory to MFRPI theory can be developed which however continues to employ the ImF premise.
Our new n-ImF theory still builds on the fundamental idea of using a trace over matrix products to define a mean-field ring-polymer potential. 
The n-ImF theory also relies on the ``ImF'' premise but it overcomes the severe break-down of MFRPI theory in the GR limit by ensuring the dominance of the second-order term in the expansion of the diabatic coupling $\Delta$ in the GR limit.
In contrast to the MFRPI theory, we observe no break-down of the \mbox{n-ImF} theory at low temperature in any regime of diabatic coupling.
In addition, we attempted to develop a related high-temperature rate expression in analogy to the Affleck rate.
It relies on the collapsed RP and predicts a rate beyond the crossover temperature.
While we could obtain reasonable results for the symmetric linear-crossing model, 
further investigation of the n-ImF high-temperature extension highlights its methodological failings.
When studying the asymmetric linear-crossing model at high temperature it becomes clear that the mean-field nature of our n-ImF rate theory forces the RP to collapse far away from the diabatic crossing point.
The method cannot therefore recover the known classical golden-rule rate.
A ring polymer in the high-friction limit also experiences unphysical behaviour.
Fortunately, the n-ImF high-temperature extension is entirely separate in derivation from our deep-tunnelling n-ImF theory.
We showed that n-ImF theory works well below crossover in comparison to benchmark results.
The instanton also allows for crucial mechanistic insights into the reaction under study, which other \textit{ad hoc} rate theories such as the interpolation formula\cite{Lawrence2019ET} cannot provide.
While we advocate for the use of our NRPI rate theory for a rigorous and efficient rate calculation, \cite{Zarotiadis2025} we think the n-ImF theory is beneficial for preliminary studies of large-scale nonadiabatic reactions.
Its simplicity, yet its ability to provide physical intuition allows one to get a useful initial perspective on the mechanism of nonadiabatic chemical reactions in the deep-tunnelling regime. %

\section*{Acknowledgements}
RAZ was supported by an Independent Postdoctoral Fellowship at the Simons Center for Computational Physical Chemistry, under a grant from the Simons Foundation (839534, MT).
The authors acknowledge financial support from the Swiss National Science Foundation through SNSF project titled ``Nonadiabatic effects in chemical reactions'' (Grant Number – 207772).
JOR thanks Prof.\ Stuart Althorpe for scientific guidance during the genesis of this work many years ago.

\appendix

%
%
%
%
%
%
%
%
%
%
%
%
%
%
%
%
%
%
%
%
%
%
%
%
%
%
%
%
%
%
%
%
%
%
%
%
%
%
%

%

%

%

%
%
%
%
%

%

%
%

%
%
%

%
%
%

%

%
%

%
%
%

%
%
%
%
%
%
%
%
%
%
%
%
%
%
%

%
%
%
%
%
%
%
%
%
%
%
%
%
%
%
%
%
%
%
%
%
%
%
%

%
\section{Derivation of the high temperature n-ImF rate expression in the GR limit}
\label{app:highT_nimf}

The proposed high-temperature n-ImF rate expression is given in Equ.~\eqref{eq:nimf_highT}.
In the adiabatic limit, it straightforwardly reduces to the version of Affleck's high-temperature rate expression given in Equ.~\eqref{eq:ad_cross_T_rate} by employing the definition of the crossover temperature $\beta_\mathrm{c} = 2 \pi/\hbar \omega_\mathrm{b}$ and the fact that $\omega_0 = \omega_\mathrm{b}$ at crossover.
Further, using that at crossover $\beta = \beta_\mathrm{c}$ we can explicitly relate both these expression to Affleck's original formula given in Equ.~\eqref{eq:Affleck_rate}.
We thus investigate in the following how our attempt at an n-ImF high-temperature extension relates to an exact expression in the GR limit.
The classical rate in the GR limit for a general system is given by\cite{GoldenGreens}
\begin{equation}
    k_\text{cl}^\text{GR} = \sqrt{\frac{2 \pi m}{\beta \hbar^2}} \frac{\Delta^2}{\hbar|\kappa_0 - \kappa_1|}\eu{-\beta V^{\ddag}}.
    \label{eq:general_1D_highT_GRrate}
\end{equation}
We can define the effective action in the golden-rule limit as
\begin{align}
    S^\text{GR} &= -\hbar \ln\Big(\frac{\Delta^2}{\hbar^2} \int_0^{\beta \hbar} \text{d} \tau'' \int_{\tau''}^{\beta \hbar} \text{d} \tau' \nonumber \\
    &\hspace{1.5cm} \eu{- S_0(0, \tau'')/\hbar - S_1(\tau'', \tau')/\hbar - S_0(\tau', \beta \hbar)/\hbar}\Big).
    \label{eq:GRaction}
\end{align}
We study the high-temperature limit of this action via the Fourier analysis with the transformations\cite{Feynman}
\begin{align}
    x_0 &= \gamma_0 \\
    x_{k_+} &= \gamma_{k_+} \sin{\nu_{k} t} = \gamma_{k_+} \sin{\frac{2 \pi k_+ t}{\beta \hbar}}\\
    x_{k_-} &= \gamma_{k_-} \cos{\nu_k t} = \gamma_{k_-} \cos{\frac{2 \pi k_- t}{\beta \hbar}},
\end{align}
with $k_+, k_- \in \mathbb{N}$.
It is helpful to first discuss the high-temperature or classical limit for the symmetric model. 
Here, it holds that the RP reduces to a single point at the barrier crossing point which is located at $x=0$.
Taking the limit $x \rightarrow 0$ is therefore assumed to be equivalent to taking the high-temperature or classical limit.
The GR action in the high-temperature limit is given by
\begin{equation}
    \lim_{x \rightarrow 0}{\eu{-S^\text{GR}(x)/\hbar}} = \beta^2 \Delta^2 \, \mathrm{e}^{-\beta V^\ddag}.
\end{equation}
The derivatives of the GR action are defined as
\begin{align}
    \lim_{x \rightarrow 0} &\frac{\partial^2 S^\text{GR}(x)}{\partial \gamma_k^2} = \nonumber \\
    &\begin{cases}
     \frac{\hbar}{3}\beta^2 \kappa^2 & \text{ $k = 0$}\\
    \frac{\hbar}{4 \beta \pi^2 k^2} \left(4 \kappa^2 \beta^3 - \frac{8 m \pi^4 k^4}{\hbar^2}\right) & \text{ $k \neq 0$}\\
    \end{cases},
\end{align}
and for the general case we define
\begin{equation*}
\lambda_0 = \frac{\partial^2 S^\text{GR}(x)}{\partial \gamma_0^2} = -\beta\hbar m\omega_0^2    \text{       and     } \omega_0 = \sqrt{\frac{- \lambda_0}{\beta\hbar m}}.    
\end{equation*}
For the symmetric case it thus holds that $\lambda_0 = \frac{\hbar}{3}\beta^2\kappa^2$ and $|\omega_0| = \sqrt{\frac{\beta\hbar m}{\lambda_0}}=\sqrt{\frac{\beta}{3 m}}\kappa$.
The crossover temperature can be inferred by setting the first eigenvalue to 0 such that
\begin{equation*}
 \lambda_{k_+=1} = \frac{\hbar}{4 \beta \pi^2}\left(4\kappa^2\beta^2 - \frac{8 m \pi^4}{\hbar^2}\right) = 0,
\end{equation*}
which can be solved as $\beta_\mathrm{c} = \left(\frac{2 m \pi^4}{\hbar^2\kappa^2}\right)^{1/3}$ in the symmetric case.
For an asymmetric linear-crossing model and still taking the limit that $x \rightarrow 0$ i.\,e.\,we assume the RP is collapsed because of high temperature, the frequency corresponding to the lowest eigenvalue of the Hessian $\left(\frac{\partial^2 S^\text{GR}}{\partial x_i \partial x_j}\right)$ is
\begin{equation}
 |\omega_0| = \sqrt{\frac{\beta}{12 m} \left(\kappa_0 - \kappa_1\right)^2}
\end{equation}
with the crossover temperature $\beta_\mathrm{c}$ correspondingly given by
\begin{equation}
 \beta_\mathrm{c} = \left( \frac{2\pi^4\beta}{3\hbar^2|\omega_0|^2}\right)^{1/3} = \left(\frac{8 m \pi^4}{\hbar^2 (\kappa_0 - \kappa_1)^2}\right)^{1/3}.
\end{equation}
Within the semiclassical approximation, the imaginary part of the partition function can be written as
\begin{align}
\text{Im}Z &= \frac{1}{2} \prod_{i=0}^{N-1} |\beta_N\hbar\omega_i|^{-1} \eu{-\beta_N U_N} \nonumber\\
    &= \frac{1}{2} |\beta_N\hbar\omega_0|^{-1} \frac{\Xi}{N} \eu{-\beta_N U_N} = \frac{1}{2} |\beta\hbar\omega_0|^{-1} \ \Xi \ \eu{-\beta_N U_N},
\end{align}
where we introduce
\begin{equation}
\Xi = \underset{N\rightarrow\infty}{\text{lim}}
    \ \prod_{i\neq0}^{N-1} \frac{\omega_i^\text{free}}{\omega_i} = N \prod_{i\neq0}^{N-1} |\beta_N \hbar \omega_i|^{-1},
\end{equation}
with $\omega_i$ as the eigenfrequencies of the RP.
Note that $\prod_{i\ne0}|\beta_N \hbar \omega^\text{free}_i| = N$ holds and $\omega_\text{free}$ corresponds to the eigenfrequencies of a free-particle RP.\cite{TuckermanBook}
The imaginary part of the free energy is then
\begin{equation}
\text{Im}F = \frac{1}{2 Z_\text{R}} \frac{\Delta^2}{\hbar|\kappa_0-\kappa_1|}\sqrt{\frac{3 m}{\beta}} \ \Xi \ \eu{-\beta V^{\ddag}}.
\label{eq:ImF_nImF}
\end{equation}
The high-temperature n-ImF rate from Equ.~\eqref{eq:nimf_highT} can therefore be simplified for a symmetric system to
\begin{align}
    k_\text{cl}^\text{n-ImF} &= \frac{2}{\hbar} \frac{\beta}{\beta_\mathrm{c}} \left(\left(\frac{2 m \pi^4}{\hbar^2\kappa^2}\right)^{1/3} \left(\frac{\beta}{3 m}\right)^{1/2} \frac{\hbar \kappa}{2\pi}\right)^{-2} \text{Im} F \nonumber \\
    &= \frac{2}{\hbar} \frac{\beta}{\beta_\mathrm{c}} \left(\frac{m \pi \hbar \kappa}{4}\right)^{-2/3}\frac{3 m}{\beta}\text{ Im} F \nonumber \\
    &= \frac{2}{\hbar} \frac{\beta}{\beta_\mathrm{c}} \left(\frac{m \pi \hbar \kappa}{4}\right)^{-2/3}\frac{3 m}{\beta} \frac{1}{2} \frac{\Delta^2}{\hbar \kappa} \sqrt{\frac{3 m}{\beta}} \frac{\Xi}{Z_\text{R}} \eu{-\beta V^{\ddag}} \nonumber \\
    &= \frac{2}{\hbar} \frac{3 \sqrt{3}}{\pi^2}\sqrt{\frac{m}{\beta}} \frac{\Xi}{Z_\text{R}}\frac{\Delta^2}{\hbar\kappa}\eu{-\beta V^{\ddag}} \nonumber \\
    &= \frac{6 \sqrt{6}}{\pi^{5/2}}\frac{\Xi}{Z_\text{R}} \sqrt{\frac{m \pi}{2\beta\hbar^2}} \frac{\Delta^2}{\hbar\kappa}\eu{-\beta V^{\ddag}} = \frac{6 \sqrt{6}}{\pi^{5/2}} \ \Xi \ k_\text{cl}^\text{GR} \nonumber \\
    &= \frac{6 \sqrt{6}}{\pi^{5/2}} \ k_\text{cl}^\text{GR}.
    \label{eq:sym_nimf_GR}
\end{align}
which deviates from the true classical rate by about 15\% given our assumptions about the classical limit of the model and the RP hold and this is within the range of error of other successful methods.\cite{Miller2003QI}
For the last line, one uses that in the classical limit
\begin{equation*}
   \lim_{\beta \rightarrow 0}{\Xi} = \lim_{m \rightarrow \infty}{\Xi} = 1.
\end{equation*}
For the more general case of an asymmetric system the definition of the derivatives accordingly changes to
\begin{align}
\lim_{\gamma_k \rightarrow 0} & \frac{\partial^2 S^\text{GR}(x)}{\partial \gamma_k^2} = \nonumber \\
&\begin{cases}
    \frac{\hbar}{12}\beta^2 (\kappa_0 - \kappa_1)^2 & \text{ $k = 0$}\\
   \frac{-\hbar}{4 \pi^2 \beta}\left(\beta^3(\kappa_0-\kappa_1)^2 - \frac{8 m \pi^4 k^4}{\hbar^2}\right) & \text{ $k \neq 0$}\\
\end{cases}.
\end{align}
Note that we keep the assumption that at high temperature the RP collapses at the crossing point $x^{\ddag}$ with $V(x^{\ddag}) = V^{\ddag}$ which for a symmetric system centred at the origin is $x=0$.
However, for a general asymmetric system it may not be the crossing point.
The rate is then obtained in the same manner as in the symmetric case as
\begin{align*}
    k_\text{cl}^\text{n-ImF}
    &= \frac{12}{\pi^2 \hbar} \beta \Delta^2 \sqrt{\frac{3 m}{\beta^3}} \frac{1}{\hbar |\kappa_0 - \kappa_1|} \eu{-\beta V^{\ddag}}.
\end{align*}
\section{Discussion of other nonadiabatic ImF rate theories}
\label{app:other_attempts}
 
Voth and coworkers developed a number of different nonadiabatic rate theories in a series of papers.\cite{Cao1995nonadiabatic, Cao1997nonadiabatic, Schwieters1998diabatic}
In each case, 
a mean-field description was used to determine the instanton pathway.
The first attempt\cite{Cao1995nonadiabatic} was based on the argument that for a semiclassical reaction path to contribute to the leading second-order term in diabatic coupling $\Delta$, the evolution on the electronic states has to be such that the electronic state is changed twice.
One can think of it as ``hopping'' from one to the other electronic state thereby picking up a factor of $\Delta$ each time.
The leading order $\Delta^2$-term hence requires two ``hops'' between diabatic states.
In the method of Ref.~\onlinecite{Cao1995nonadiabatic}, the beads which are forced to hop are chosen such that they split the path into two equal halves.
In this way, they suppress the contribution from the zero-hop term and the leading-order term with respect to the coupling strength is constructed to be of second order.
This guarantees that the rate will follow the $\Delta^2$-dependence of Fermi's GR in the GR limit.
However, the hops are in general not located optimally for an asymmetric system. %
Additionally, this approach breaks the fundamental permutational symmetry of the ring polymer.
Breaking the permutational symmetry introduces significant errors in the adiabatic limit, as was pointed out in subsequent publications by Schwieters and Voth.\cite{Schwieters1998diabatic, Schwieters1999diabatic}
In order to improve upon this, Schwieters and Voth\cite{Schwieters1998diabatic} presented the MFRPI approach discussed in this work.
We have, however, shown that this approach breaks down in the golden-rule limit.

Our n-ImF theory goes beyond both these previous approaches by ensuring that at least two hops are made so as to describe the GR limit, while maintaining permutational invariance so as to reduce to the BO instanton.

\section{Classical Holstein theory}
\label{app:Holstein}
It has been proposed to calculate the nonadiabatic rate
using \cite{Nitzan,PetersBook}
\begin{equation}
k_\text{cl} = \frac{1}{Z_\text{R}} \int_0^{\infty} \text{d$v$ } T_\text{H}(v) \frac{m}{2\pi\hbar}\,\eu{-\beta m v^2/2}\,\eu{-\beta(V^\ddag-\Delta)},
\end{equation}
where $v$ is the velocity
and Holstein's transmission probability is \cite{Holstein1959b}
\begin{equation}
T_\text{H}(v) = \frac{2 T_\text{LZ}(v)}{1 + T_\text{LZ}(v)}.    
\end{equation}
The Landau--Zener transition probability to move from one to the other diabat is given by\cite{Landau1932LZ, Zener1932LZ}
\begin{equation}
T_\text{LZ}(v) = 1- \exp\left(-\frac{2 \pi\Delta^2}{|\kappa_0 - \kappa_1| v \hbar}\right).
\end{equation}
where $\kappa_0$ and $\kappa_1$ are the slopes of the two diabats at the crossing point.
\section{Interpolation formula}
\label{app:IF}

The idea of interpolation to overcome the limits of weak and strong diabatic coupling has already been investigated in a number of ways.\cite{Zusman1980, Gladkikh2005, Garg1985spinboson, Hynes1986, RipsJortner1987, SparpaglioneMukamel1988}
The interpolation formula of Lawrence et al.\cite{Lawrence2019ET} revives these earlier ideas and it thereby aims to bridge the nonadiabatic gap in quantum rate theories.
The IF approach requires three separate rate calculations according to its key equation
\begin{equation}
   k_\text{IF}(\Delta) = \frac{k_\text{GR}(\Delta) k_\text{BO}(\Delta) }{k_\text{GR}(\Delta) + k_\text{BO}(\Delta = 0)},
\end{equation}
namely a Wolynes rate calculation, and two RPMD rate calculations at $\Delta=0$ and at the coupling strength of the system.
The limiting rate theories could of course easily be replaced by their instanton counterparts, an idea which was already mentioned in Ref.~\onlinecite{Trenins2020}.
A key advantage of the IF method is that it strictly recovers the limiting theories i.e., it tends to $k_\text{GR}$ in the GR limit and returns $k_\text{BO}$ for larger diabatic couplings.\cite{Lawrence2019ET}
Nonetheless, the IF approach remains fundamentally \textit{ad hoc}.
Additionally, it inherits the errors of the limiting rate theories\cite{RPInst,Fe2Fe3, Vaillant2019QInst}. %
The most substantial criticism of the IF approach relates to its inability to capture the true nature of the reaction mechanism.
In the IF approach the nonadiabatic rate is obtained by interpolation between two limiting rate theories and in general neither of the rate mechanisms of these limiting cases can be expected to accurately depict the true mechanism of the nonadiabatic reaction.
Ultimately, if the fundamental nature of the reaction mechanism differs from the limiting cases, the IF approach cannot be expected to accurately predict the rate coefficient.

\bibliography{references,references_rhi}
\end{document}